%% file: combined.tex
\documentclass[%preprint, linenumbers,
 amsmath,amssymb,
 aps, physrev,
prl,
twocolumn
]{revtex4-2}
\usepackage{xcolor}
\usepackage{cancel}
\usepackage{braket}
\usepackage{graphicx}% Include figure files
\usepackage{dcolumn}% Align table columns on decimal point
\usepackage{bm}% bold math
\begin{document}

% ================= MAIN PAPER =================

\input{main}

% ================= SUPPLEMENT =================
\clearpage
\appendix

\input{supplement}

\end{document}

%% file: main.tex
%\preprint{APS/123-QED}

\title{\textbf{Effect of Electron Correlation on the Integer Quantum Hall Effect}
}% 

\author{Daniel Staros\textsuperscript{1}}
 \email{Contact author: dstaros@lanl.gov}
\author{Christopher Lane\textsuperscript{1}}%
\author{Roxanne Tutchton\textsuperscript{1}}%
\author{Jian-Xin Zhu\textsuperscript{1,\textdagger}}
\affiliation{%
 \mbox{\textsuperscript{1}Theoretical Division, Los Alamos National Laboratory, Los Alamos, New Mexico 87545, USA}\\
\mbox{%
\textsuperscript{\textdagger}Center for Integrated Nanotechnologies, Los Alamos National Laboratory, Los Alamos, New Mexico, 87545, USA}\\
}%

%\date{\today}

\begin{abstract}
We numerically investigate the effect of electron correlation on the integer quantum Hall effect in a square lattice. Increasing the correlation strength via the effective onsite repulsion parameter $U$ degrades the quantization of $\nu = 1$ transverse conductance due to the interplay of correlation and the external magnetic field, which together induce periodic modulations in renormalized hopping parameters and site energies. Overall, this work demonstrates that the strength of electron correlation can significantly impact conductivity in the integer quantum Hall regime.
\end{abstract}

\maketitle

%\tableofcontents

%\section{\label{sec:intro}Introduction}

\textit{Introduction.}---The integer quantum Hall effect (IQHE) is a cornerstone of modern condensed matter theory \cite{Klitzing_prl1980, Prange_1989, Janssen_1994, Bernevig_2012}. Fundamentally, it serves as a first example of topological quantum phase. The robust integer quantization of transverse conductivity, which defines the IQHE has led to multiple technological innovations including metrological standards for electrical resistance and IQHE-based interferometers for probing anyonic quasiparticle statistics~\cite{Hartland_met1992, Dutta_jap2012, Werkmeister_natcomm2024}. The robustness of the IQHE stems from the topological protection afforded by highly degenerate cyclotron orbits, or Landau levels (LLs), against external perturbations. One type of perturbation is site disorder, an increase in which was previously shown to degrade the integer conductance in a noninteracting lattice model \cite{Dutta_jap2012,Thomas_prb2025}. Although its effects could also be considered a perturbation, electron correlation is presumed to play a minor role in the IQHE when the disorder effect is dominant, in contrast to the fractional quantum Hall effect \cite{Stormer_rmp2025} where correlation plays a major role. Accordingly, there has been little motivation to investigate the effect of correlation on microscopic models of the IQHE. 

Very recently, experimental evidence of an interaction-driven breakdown of the IQHE in GaAs/AlGaAs quantum wells was reported \cite{West_arxiv2025}. Specifically, a transport instability for the $\nu=1$ LL was identified, and a correlation-induced asymmetry in the dissipation current for filling just above and below $\nu=1$ provided evidence that the instability is dominated by electron-electron interactions. This is a striking demonstration that correlation can have an important impact on the IQHE. However, a fundamental understanding of the origin of such phenomena remains missing. In fact, the effect of correlation on the IQHE has been virtually unexplored in general, leaving a gap in understanding that could help bridge the single-particle-like nature of the IQHE with the many-body nature of the fractional quantum Hall effect \cite{West_arxiv2025}.

In this Letter, we present a simplified study of the effect of electron correlation on the IQHE. Within the Gutzwiller approximation, we elucidate how electron correlation renormalizes the bands of electrons on a square lattice subject to a strong, perpendicular magnetic field. Our band renormalization accounts for the fact that the translation invariance of the square lattice is modified by the external magnetic field~\cite{Bernevig_2012}. Remarkably, we find a key similarity to the recent experiment of Ref. \cite{West_arxiv2025}, where correlation degrades the quantization of our $\nu = 1$ LL. We identify the correlation-induced closing of the mobility gap between the $\nu = 1$ and $\nu = 2$ LLs as the explanation for this effect, and show that band bending, ultimately due to spatial modulations of the local electronic structure, underlies this gap closing. Our work thus establishes that correlation alone can degrade the IQHE in spatially unrestricted Gutzwiller-renormalized lattice models, and outlines a clear electronic mechanism. More broadly, it provides a conceptual link between noncorrelated (integer) and correlated (integer-degraded) quantum Hall effects from a real-space perspective, which allows for the spatial modulation.

\begin{figure}[b!]
\includegraphics[width=\linewidth]{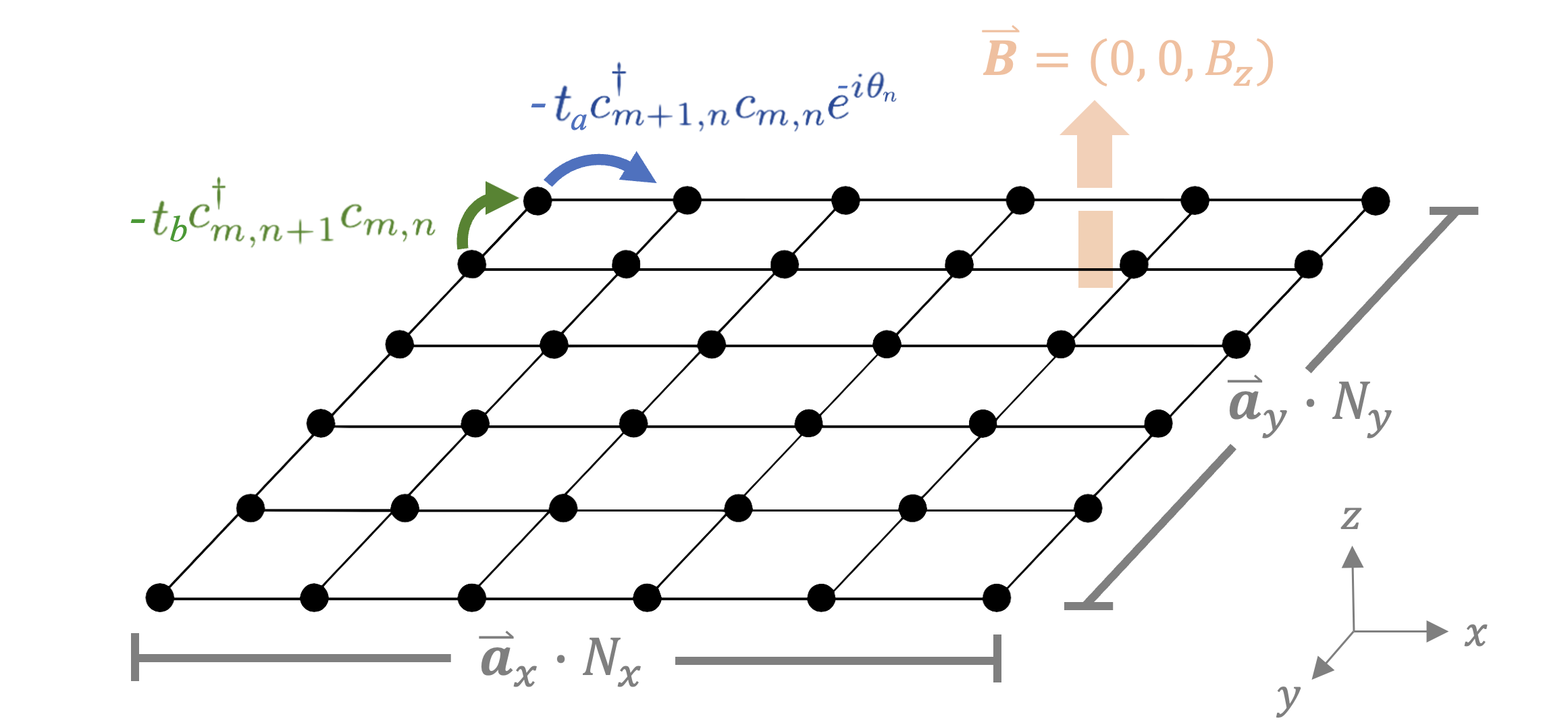}
%\vspace*{-15mm}
\caption{\label{fig:lattice} Schematic of the magnetized square lattice model illustrating nearest neighbor hopping parameterized by $t_a$ and $t_b$, lattice dimensions and perpendicular magnetic field $B$.}
\end{figure}

%\section{\label{sec:methods}Lattice model}
\textit{Theoretical model and methods.}---We consider a Hubbard model defined on a two-dimensional (2D) square lattice with a strong magnetic field applied perpendicular the lattice plane, as schematically shown in Fig.~\ref{fig:lattice}. Within a Landau gauge, the model  Hamiltonian can be written in real space as
\begin{gather}
    \hat{H} = \sum_{m,n,s} \biggl{\{} \epsilon_{m,n,s} c_{m,n,s}^{\dagger} c_{m,n,s} - \biggl{(}t e^{-i2\pi\phi n}c_{m+1,n,s}^{\dagger}c_{m,n,s}  \nonumber \\ 
    + t c_{m,n+1,s}^{\dagger}c_{m,n,s}+\textnormal{H.c.}\biggr{)}\biggr{\}} + \sum_{m,n} U n_{m,n,\uparrow} n_{m,n,\downarrow}\;.
    \label{eqn:hub_ham_r}
\end{gather}
\noindent
Here the operators $c_{m,n,s}^{\dagger}$ ($c_{m,n,s}$) create (annihilate) an electron at site $\textbf{r}_{m,n}=ma \hat{x}+na \hat{y}$ for integers $m$ and $n$ such that $ 0 \leq m < N_x$ and $ 0 \leq n < N_y$ (we choose $a=1$) with periodic boundary conditions. Local onsite energies are denoted by $\epsilon_{m,n,s}$, the physical quantity $t$ is the hopping integral along $\hat{x}$ and $\hat{y}$, and $s \in \{\uparrow,\downarrow\}$ indexes the spin. The perpendicular magnetic field $\mathbf{B}=(0,0,B_z)$ has vector potential $\mathbf{A}=(-yB_z,0,0)$ where $\phi=1/q$ defines the corresponding magnetic flux per plaquette and $q$ is commensurate with $N_y$.

\begin{figure*}[t!]
\includegraphics[width=6.5in]{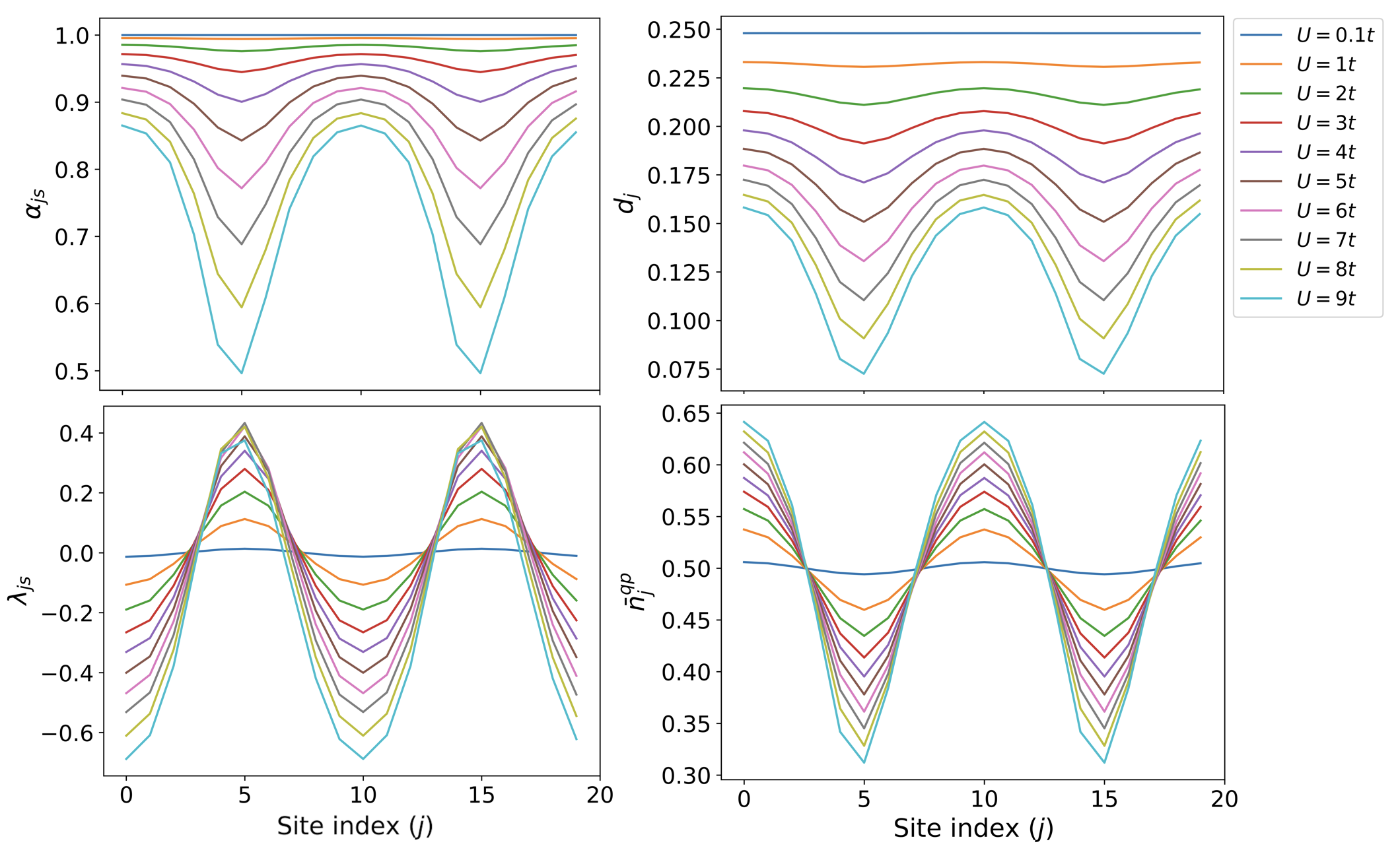}
%\vspace*{-15mm}
\caption{Optimal parameters as a function of site index and $U$ ($q=20$); (top left) hopping renormalization $\alpha_{j,s}$, (top right) double occupancy $d_j$, (bottom left) site renormalization energies $\lambda_{j,s}$, and (bottom right) quasiparticle occupations $\bar{n}_{j,s}^{qp}$.}
\label{fig:params}
\end{figure*}

Under the Gutzwiller wavefunction approximation~\cite{Gutzwiller_pr1965, Julien_2006, Zhu_prl2012, Nandy_prb2024}, the Hamiltonian of Eq.~(\ref{eqn:hub_ham_r}) can be partitioned into $q$ sectors along $k_y$ due to the magnetic field,
\begin{gather}
    H = \sum_s \sum_{k_x=-\pi}^{\pi} \sum_{k_y=-\pi/q}^{\pi/q} \hat{h}_s(k_x,k_y) + N_{k_x}N_{k_y}\sum_{j=0}^{q-1} U d_j
\label{eqn:gutz_ham_mix}
\end{gather}
\noindent
where $N_{k_x}=N_x$ and $N_{k_y} = N_y/q$ of the corresponding real-space lattice of size $N_x \times N_y$, and
\begin{gather}
    \hat{h}_s(k_x,k_y) = \sum_{j=0}^{q-1}
    \left[(\epsilon_{j,s} + \lambda_{j,s}) - 2t \alpha_{j,s}\cos(k_x+2\pi \phi j) \right] \times \nonumber \\  c_{k_x,k_y,j,s}^{\dagger}  c_{k_x,k_y,j,s} - t \sqrt{\alpha_{j,s}} \left(
     e^{-ik_y} \sqrt{\alpha_{j+1,s}} c_{k_x,k_y,j+1,s}^{\dagger}  \right. \\ \left. + e^{ik_y} \sqrt{\alpha_{j-1,s}}c_{k_x,k_y,j-1,s}^{\dagger}\right) c_{k_x,k_y,j,s}\;. \nonumber
\label{eqn:gutz_ham_plaq}
\end{gather}
\noindent
Here, $k_x$ and $k_y$ are the $x$- and $y$-components of the reciprocal lattice vector describing periodicity of the magnetized unit cell, and $j$ indexes the $j^{th}$ plaquette within a unit cell at a given reciprocal lattice point. The local electronic structure of our solutions is governed by: (1) the Gutzwiller renormalization parameters $\alpha_{j,s}$, (2) the double occupancy parameters $d_j$, and (3) the site renormalization energies $\lambda_{j,s}$. Hopping parameters are scaled by Gutzwiller renormalization factors $\sqrt{\alpha_{j,s}}$ defined by
\begin{gather}
    \sqrt{\alpha_{j,s}} = \left[\frac{\left(\bar{n}_{j,s}-d_{j}\right)\left(1-\sum_{s}\bar{n}_{j,s}+d_{j}\right)}{\bar{n}_{j,s}(1-\bar{n}_{j,s})}\right]^{1/2} \nonumber \\ +\left[\frac{d_{j}\left(\bar{n}_{j,\bar{s}}-d_{j}\right)}{\bar{n}_{j,s}(1-\bar{n}_{j,s})}\right]^{1/2}, 
\label{eqn:alpha_def}
\end{gather}
\noindent
where stronger correlation (larger values of $U$) corresponds to smaller values of $\sqrt{\alpha}$, thereby suppressing hopping between sites. We self-consistently solve for the values of $\{\alpha_{j,s}\}$, $\{d_j\}$ and $\{\lambda_{j,s}\}$ that minimize the expectation value of the Hamiltonian in Eq.~(\ref{eqn:gutz_ham_mix}) subject to the self-consistency conditions on $\lambda_{j,s}$ and $d_j$ that are provided in the SI.

%\section{\label{sec:elec_structure} Local electronic structure}
\textit{Electronic structure.}---To investigate the effect of correlation on the electronic structure of the magnetized square lattice, we test our model over correlation strengths ranging from $U = 0.1t$ to $U=9t$. The real-space and $k$-space grids underlying these calculations correspond to $N_x = N_y = 20$, which are sufficiently converged to resolve quantization of the first three Landau levels in the noninteracting limit. We choose the hopping integral $t = 1$ to ensure that any anisotropy in the optimized Gutzwiller parameters is solely due to the interplay between the correlation and magnetic orbital effects, and consider the smallest value of the magnetic field which is consistent with the given lattice dimensions, $B = 2\pi/q$ with $q = 20$. All Gutzwiller optimizations are performed at half filling. As a rule of thumb, solutions of a Gutzwiller-renormalized lattice model are best behaved over a range of $U$ values corresponding to twice the total bandwidth. Our model converges up to $U=8t$, converges with self-consistent mixing at $U=9t$, but is significantly more challenging to converge for greater values of $U$. This is consistent with our bandwidth of $\sim 8t$.

\begin{figure*}[t!]
\includegraphics[width=6.5in]{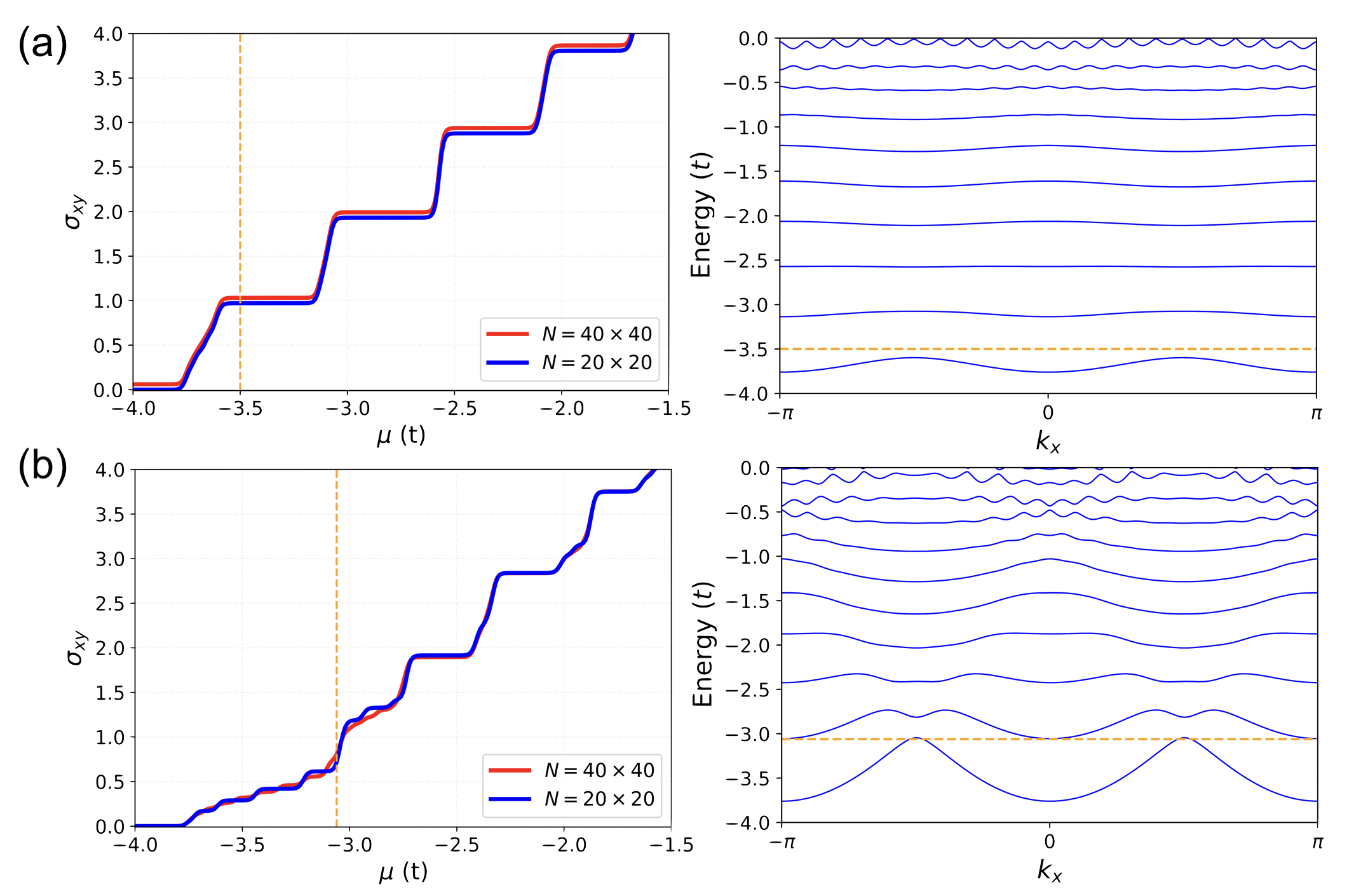}
%\vspace*{-15mm}
\caption{Plots of transverse conductivity in units of $e^2/h$ vs. chemical potential $\mu$ (left) and corresponding $k_x$ ($k_y=0$) band dispersion (right) optimized for (a) $U=1t$ and (b) $U=5t$. The red line at top left is vertically offset by $+0.06$ for clarity.}
\label{fig:sigma_bands}
\end{figure*}

Figure \ref{fig:params} shows the solutions for $\alpha$, $d$, and $\lambda$ obtained for various values of the effective Coulomb repulsion parameter $U$, as well as the corresponding quasiparticle occupations $\bar{n}_{j,s}$. All of the parameters in our model exhibit periodicity as a function of site index $N_{y_j}$ which corresponds to the magnitude of the Peierls phase that defines the magnetic flux per plaquette. Larger values of $\alpha$ and $d$ indicate rows along which hopping is more favorable and sites are more likely to be doubly occupied, while smaller values indicate the opposite. Additionally, the behavior of the site renormalization energies $\lambda_{j,s}$ is consistent with that of the hopping renormalization parameters; sites with lower energies are less likely to be doubly occupied than those with higher energies. Lastly, we quantify the effect of band renormalization on the charge distribution, seeing that rows with lower site energies have greater average occupations than those with higher site energies. 

Overall, these results demonstrate that the interplay between an applied magnetic field and electron correlation in a square lattice results in the spatial modulation of site renormalization energies, hopping renormalization parameters, and thus the quasiparticle charge distribution within the Gutzwiller approximation. Specifically, increasing the effective onsite repulsion $U$ results in larger intracell electronic structure modulation, quantified by the difference between the smallest and largest row-wise parameter values, indirectly suggesting that $U$ can be used as a tuning knob to modify the transport properties of the system.

%\section{\label{sec:conductivity} Conductivity with correlation}
\textit{Effects of correlation on conductivity.}---In order to directly see how electrical transport may be affected by the correlation and field-induced modulation of the local electronic structure in the square lattice, we calculate the transverse conductivity for the cases of weak ($U=1t$) and moderate ($U=5t$) electron correlation. Our calculations are performed using the Kubo formula for transverse conductivity $\sigma_{xy}$ as given by
\begin{gather}
    \sigma_{xy}(\mu) = \frac{ie^2 \hbar}{N} \sum_{a} \sum_{b \neq a} \left(f_{\epsilon_a}-f_{\epsilon_b}\right) \frac{\braket{a|\hat{v}_x|b}\braket{b|\hat{v}_y|a}}{(\epsilon_a-\epsilon_b)^2+\delta^2}
    \label{eqn:kubo}
\end{gather}
\noindent
where $\ket{a}$ and $\ket{b}$ are eigenvectors with corresponding eigenenergies $\epsilon_a$ and $\epsilon_b$, which belong to the set $\left\lbrace \ket{\gamma}, \epsilon_{\gamma} \right\rbrace$ obtained by diagonalizing the real-space tight-binding Hamiltonian, $f_{\epsilon_{\gamma}} = \left[1+ \textnormal{exp}\left(\frac{\epsilon_{\gamma}-\mu}{k_B T}\right)\right]^{-1}$ is the Fermi-Dirac distribution as a function of $\mu$ for a given energy $\epsilon_{\gamma}$, and $\delta$ is the level broadening parameter \cite{Dutta_jap2012}. Similarly to a previous calculation of $\sigma_{xy}$ in a non-interacting square lattice model, we adopt natural units such that $e=c=h=1$ and fixed $k_BT=0.01$ \cite{Dutta_jap2012}. We also set $d=0.01$ to ensure that the level broadening is negligible compared to bandwidths. The derivation and expressions for the velocity operators $\hat{v}_x$ and $\hat{v}_y$ for our Gutzwiller-renormalized lattice model are provided in the SI. We note that the conductivity is evaluated in real space under periodic boundary conditions.

Figure \ref{fig:sigma_bands} presents the transverse conductivity $\sigma_{xy}$ and the corresponding band structure of our model for both correlation strengths. In the weakly correlated case (Figure \ref{fig:sigma_bands}a), there are well-defined integer steps in $\sigma_{xy}$ which correspond to the energies of the isolated Landau levels visible in the corresponding band structure. In contrast, moderate correlation (Figure \ref{fig:sigma_bands}b) degrades the integer quantization of the $\nu=1$ Landau level. Nearly identical agreement between $\sigma_{xy}$ for a $20 \times 20$ and $40 \times 40$ lattice demonstrates that finite-size effects are negligible. The renormalized band structures suggest instead that the quantization degradation of the $\nu=1$ LL stems from a correlation-induced closing of the mobility gap between LLs $\nu=1$ and $\nu=2$. Specifically, correlation causes the bands corresponding to the first two Landau levels to bend until they are no longer isolated in energy space, as indicated by the horizontal line in Figure \ref{fig:sigma_bands}. In other words, the correlation-induced modulation in $\alpha_{j,s}$ and $\lambda_{j,s}$ results in the closing of the first mobility gap. Although band dispersions near the other Landau levels are also bent when $U=5t$, the effect is not large enough to close any of the remaining mobility gaps, explaining their robust integer quantization. We note that our results contrast with the effect of random site disorder in the IQHE of a noninteracting lattice model, the increase of which degrades the quantization of all Landau levels \cite{Dutta_jap2012}. This is because the introduction of random disorder removes local translational invariance, resulting in the collapse of extended conducting states to localized insulating states \cite{Huckestein_rmp1995}. In contrast, our model maintains the translation symmetry of the magnetized square lattice, so our solutions remain extended (Bloch-like) even in the presence of such intracell parameter modulations \cite{Bernevig_2012}.

Lastly, we calculate renormalized band structures in the absence of site renormalization parameters $\lambda_{j,s}$ to identify which parameter dominates the observed degradation of the $\nu=1$ LL. These results (provided in the SI) show that the band bending originates from the modulation of the hopping renormalization parameters $\alpha_{j,s}$, but that the magnitude of the bending is additionally dependent on $\lambda_{j,s}$. Specifically, the renormalized bands for $U=5t$ are clearly bent when all $\lambda_{j,s}=0$, but the effect is not large enough to close the mobility gap between the first and second LLs. In other words, the  hopping renormalization alone does impact the size of the mobility gap, but the degradation of the first Landau level requires that effective charge reordering, quantified by the site energies, also be considered.  

%\section{\label{sec:discussion}Conclusion}
\textit{Conclusion.}---In summary, we find that strong electron correlation can significantly alter the integer quantum Hall effect in a 2D square lattice. Specifically, increased Coulomb repulsion results in the degradation of the integer quantization of the transverse conductivity of the first Landau level due to the closing of the mobility gap between LLs $\nu=1$ and $\nu=2$. This observation is explained by band bending that results from the joint effect of correlation and topology on the electronic structure of the magnetized square lattice. The spatial modulation of hopping renormalization parameters enables the band bending, while the additional bending from the charge reorganization results in the closing of the lowest-lying mobility gap. Our results thus share a key similarity with recent experimental results, in which a correlation-driven breakdown of the $\nu=1$ Landau level was proposed. Indirectly, our work helps bridge the conceptual gap between integer and fractional quantum Hall phases by directly considering the effect of strong correlation on integer quantum Hall states. \\

\section{\label{sec:acknowledgements}Acknowledgments}
The authors thank Jean-Pierre Julien for helpful discussions. This work was carried out under the auspices of the U.S. Department of Energy (DOE) National Nuclear Security Administration under Contract No. 89233218CNA000001, and supported by the LANL LDRD Program, project \#20230042DR, ``Implications of Excess Electronic Entropy for the Phase Diagram of Plutonium”, and was performed, in part, at the Center for Integrated Nanotechnologies, an Office of Science User Facility operated for the U.S. Department of Energy (DOE) Office of Science in partnership with the LANL Institutional Computing Program for computational resources.

\bibliography{main}

%\end{document}

%% file: supplement.tex
\onecolumngrid

\section*{\label{sec:supp_inf}Supplementary Information for ``Effect of Electron Correlation on the Integer Quantum Hall Effect"}

We provide here the additional information necessary to replicate our study, but which does not comprise the key scientific findings. Specifically, we first provide the derivation of the self-consistency conditions in real space, since these have not been provided in the context of a model like ours. Then, we supply the derivation of the mixed-space representation of our lattice model, since, to the best of our knowledge, it has not previously been written down in the way we present it in the manuscript. Third, we explicitly derive the form of the velocity operators that enter the Gutzwiller-renormalized Kubo formula for conductivity. Lastly, we provide additional results clarifying how the site renormalization plays a key role in the mobility gap closing observed in the main manuscript.

\appendix
\section{\label{sec:conditions}Derivation of Self-Consistency Conditions}

In our chosen gauge, the Gutzwiller-renormalized square lattice (with lattice vector $a=1$) in the presence of an external field is
\setcounter{equation}{0}
\begin{gather}
    H =  \sum_{m,n,s} \left(\epsilon_{m,n,s} +\lambda_{m,n,s}\right) c_{m,n,s}^{\dagger} c_{m,n,s} -t_a\sqrt{\alpha}_{m+1,n,s}\sqrt{\alpha}_{m,n,s}(e^{-i\theta_n}c_{m+1,n,s}^{\dagger}c_{m,n,s}+\textnormal{H.c.}) \nonumber \\ - t_b \sqrt{\alpha}_{m,n+1,s}\sqrt{\alpha}_{m,n,s}(c_{m,n+1,s}^{\dagger}c_{m,n,s}+\textnormal{H.c.}) + \sum_{m,n} U d_{m,n},
\label{eqn:ham}
\end{gather}
\noindent
and its expectation value is
\begin{gather}
    \braket{H} =  \sum_{m,n,s} \left(\epsilon_{m,n,s} +\lambda_{m,n,s}\right) \braket{c_{m,n,s}^{\dagger} c_{m,n,s}} -t_a\sqrt{\alpha}_{m+1,n,s}\sqrt{\alpha}_{m,n,s}(e^{-i\theta_n}\braket{c_{m+1,n,s}^{\dagger}c_{m,n,s}}+\textnormal{c.c.}) \nonumber \\ - t_b \sqrt{\alpha}_{m,n+1,s}\sqrt{\alpha}_{m,n,s}(\braket{c_{m,n+1,s}^{\dagger}c_{m,n,s}}+\textnormal{c.c.}) + \sum_{m,n} U d_{m,n}.
\label{eqn:exp}
\end{gather}
\noindent
\noindent
The derivative of the expectation value with respect to $\bar{n}_{m,n,\uparrow}$ is then
\begin{gather*}
    \frac{\partial \braket{H}_{m,n}}{\partial \bar{n}_{m,n,\uparrow}} = \frac{\partial}{\partial \bar{n}_{m,n,\uparrow}} \sum_{s} \left[ \left(\epsilon_{m,n,s} +\lambda_{m,n,s}\right) \bar{n}_{m,n,s} -t_a\sqrt{\alpha}_{m+1,n,s}\sqrt{\alpha}_{m,n,s}(e^{-i\theta_n}\braket{c_{m+1,n,s}^{\dagger}c_{m,n,s}}+\textnormal{c.c.}) \nonumber \right.\\ \left. - t_b \sqrt{\alpha}_{m,n+1,s}\sqrt{\alpha}_{m,n,s}(\braket{c_{m,n+1,s}^{\dagger}c_{m,n,s}}+\textnormal{c.c.}) \right] \\
    = \epsilon_{m,n,\uparrow} + \lambda_{m,n,\uparrow} \\ - \left[t_a\sqrt{\alpha}_{m+1,n,\uparrow}\left(\frac{\partial \sqrt{\alpha}_{m,n,\uparrow}}{\partial \bar{n}_{m,n,\uparrow}}\right) e^{-i\theta_n}\braket{c_{m+1,n,\uparrow}^{\dagger}c_{m,n,\uparrow}} +  t_b \sqrt{\alpha}_{m,n+1,\uparrow}\left(\frac{\partial \sqrt{\alpha}_{m,n,\uparrow}}{\partial \bar{n}_{m,n,\uparrow}}\right)\braket{c_{m,n+1,\uparrow}^{\dagger}c_{m,n,\uparrow}} \right. \\ \left.
    + t_a\sqrt{\alpha}_{m+1,n,\downarrow}\left(\frac{\partial \sqrt{\alpha}_{m,n,\downarrow}}{\partial \bar{n}_{m,n,\uparrow}}\right) e^{-i\theta_n}\braket{c_{m+1,n,\downarrow}^{\dagger}c_{m,n,\downarrow}} + t_b \sqrt{\alpha}_{m,n+1,\downarrow}\left(\frac{\partial \sqrt{\alpha}_{m,n,\downarrow}}{\partial \bar{n}_{m,n,\uparrow}}\right)\braket{c_{m,n+1,\downarrow}^{\dagger}c_{m,n,\downarrow}} \right] + \textnormal{c.c.} \\
    = \epsilon_{m,n,\uparrow} + \lambda_{m,n,\uparrow} - \left[\frac{\partial \sqrt{\alpha}_{m,n,\uparrow}}{\partial \bar{n}_{m,n,\uparrow}} \left(t_a\sqrt{\alpha}_{m+1,n,\uparrow} e^{-i2\pi \phi n}\braket{c_{m+1,n,\uparrow}^{\dagger}c_{m,n,\uparrow}} + t_b \sqrt{\alpha}_{m,n+1,\uparrow}\braket{c_{m,n+1,\uparrow}^{\dagger}c_{m,n,\uparrow}} \right) \right. \\ \left.
    + \frac{\partial \sqrt{\alpha}_{m,n,\downarrow}}{\partial \bar{n}_{m,n,\uparrow}} \left(t_a\sqrt{\alpha}_{m+1,n,\downarrow} e^{-i2\pi \phi n}\braket{c_{m+1,n,\downarrow}^{\dagger}c_{m,n,\downarrow}} + t_b \sqrt{\alpha}_{m,n+1,\downarrow}\braket{c_{m,n+1,\downarrow}^{\dagger}c_{m,n,\downarrow}} \right) \right] + \textnormal{c.c.}
\end{gather*}
\noindent
Thus, the corresponding minimum of $\braket{H}_{m,n}$ occurs where the above LHS is equal to 0, or
\begin{gather*}
    \lambda_{m,n,\uparrow} = -\epsilon_{m,n,\uparrow}  +\left[\frac{\partial \sqrt{\alpha}_{m,n,\uparrow}}{\partial \bar{n}_{m,n,\uparrow}} \left(t_a\sqrt{\alpha}_{m+1,n,\uparrow} e^{-i2\pi \phi n}\braket{c_{m+1,n,\uparrow}^{\dagger}c_{m,n,\uparrow}} + t_b \sqrt{\alpha}_{m,n+1,\uparrow}\braket{c_{m,n+1,\uparrow}^{\dagger}c_{m,n,\uparrow}} \right) \right. \\ \left.
    + \frac{\partial \sqrt{\alpha}_{m,n,\downarrow}}{\partial \bar{n}_{m,n,\uparrow}} \left(t_a\sqrt{\alpha}_{m+1,n,\downarrow} e^{-i2\pi \phi n}\braket{c_{m+1,n,\downarrow}^{\dagger}c_{m,n,\downarrow}} + t_b \sqrt{\alpha}_{m,n+1,\downarrow}\braket{c_{m,n+1,\downarrow}^{\dagger}c_{m,n,\downarrow}} \right) \right] + \textnormal{c.c.}
\end{gather*}
\noindent
Generally, this reflects the constraint that 
\begin{gather*}
    \lambda_{m,n,s} = -\epsilon_{m,n,s}  +\left[\frac{\partial \sqrt{\alpha}_{m,n,s}}{\partial \bar{n}_{m,n,s}} \left(t_a\sqrt{\alpha}_{m+1,n,s} e^{-i2\pi \phi n}\braket{c_{m+1,n,s}^{\dagger}c_{m,n,s}} + t_b \sqrt{\alpha}_{m,n+1,s}\braket{c_{m,n+1,s}^{\dagger}c_{m,n,s}} \right) \right. \\ \left.
    + \frac{\partial \sqrt{\alpha}_{m,n,\bar{s}}}{\partial \bar{n}_{m,n,s}} \left(t_a\sqrt{\alpha}_{m+1,n,\bar{s}} e^{-i2\pi \phi n}\braket{c_{m+1,n,\bar{s}}^{\dagger}c_{m,n,\bar{s}}} + t_b \sqrt{\alpha}_{m,n+1,\bar{s}}\braket{c_{m,n+1,\bar{s}}^{\dagger}c_{m,n,\bar{s}}} \right) \right] + \textnormal{c.c.}
    \label{eqn:lambda_def}
\end{gather*}
\noindent
for each spin site \{$m,n,s$\}.

Similarly, the requirement that $\frac{\partial \braket{H}_{m,n}}{\partial d_{m,n}} = 0$ yields the second set of self-consistency conditions given by
\begin{gather}
    U = \sum_s \left(\frac{\partial \sqrt{\alpha_{m,n,s}}}{\partial d_{m,n}}\right)\left[t_a \sqrt{\alpha_{m+1,n,s}}e^{-i\theta_n}\braket{c_{m+1,n,s}^{\dagger}c_{m,n,s}} + t_b \sqrt{\alpha_{m,n+1,s}}\braket{c_{m,n+1,s}^{\dagger}c_{m,n,s}}+\textnormal{c.c.} \right].
    \label{eqn:d_def}
\end{gather}

\section{Derivation of Mixed-Space Model}

As for the previous derivation, we start with the real-space Hamiltonian of a Gutzwiller renormalized square lattice (with lattice vector $a=1$) subject to an external magnetic field,

\begin{gather}
    H = H_{kin}(\textbf{R}) + H_{pot}(\textbf{R}) \nonumber \\  =  \sum_{m,n,s} \left(\epsilon_{m,n,s} +\lambda_{m,n,s}\right) c_{m,n,s}^{\dagger} c_{m,n,s} -t_a\sqrt{\alpha_{m+1,n,s}}\sqrt{\alpha_{m,n,s}}(e^{-i2\pi \phi n}c_{m+1,n,s}^{\dagger}c_{m,n,s}+\textnormal{H.c.}) \nonumber \\ - t_b \sqrt{\alpha_{m,n+1,s}}\sqrt{\alpha_{m,n,s}}(c_{m,n+1,s}^{\dagger}c_{m,n,s}+\textnormal{H.c.}) + \sum_{m,n} U d_{m,n}.
\label{eqn:gutz_ham_r}
\end{gather}

Dropping the spin index, one may write the kinetic part of Equation \ref{eqn:ham} as
\begin{gather*}
    H_{kin}(\textbf{R}) = H_x + H_y \\
    H_x = -t_a \sum_{m,n} \alpha_n \left[e^{-i2\pi \phi n}c_{m+1,n}^{\dagger}c_{m,n} + \textnormal{h.c.}\right] \; \; \; \; \; \; \; \; \; \; \; H_y = -t_b \sum_{m,n} \sqrt{\alpha}_{n+1}\sqrt{\alpha}_n c_{m,n+1}^{\dagger} c_{m,n} 
\end{gather*}
\noindent
The translation symmetry of the magnetized square lattice requires that $n = lq+j$, where $l$ indexes the unit cell along $y$ and $j$ indexes the plaquette within the unit cell, meaning that
\begin{gather*}
    c_{m,n}^{\dagger}=c_{m,l,j}^{\dagger} = \frac{1}{\sqrt{N_x}}\frac{1}{\sqrt{N_y/q}} \sum_{k_x=-\pi}^{\pi} \sum_{k_y=-\pi/q}^{\pi/q} e^{-i[mk_x+(lq+j)k_y]} c_{k_x,k_y,j}^{\dagger}.
\end{gather*}
\noindent
Noting that $e^{i\theta_n} = e^{i2\pi\phi n} = e^{i2\pi \phi j} e^{i2\pi \phi lq} = e^{i2\pi \phi j}$ and $\alpha_n = \alpha_{j+lq} = \alpha_j$ by symmetry,
\begin{gather*}
    \sum_{m,n}\alpha_n e^{-i2\pi \phi n}c_{m+1,n}^{\dagger}c_{m,n} = \frac{q}{N_x N_y} \sum_{k_x,k_y}\sum_{k_x',k_y'} e^{-ik_x}
    \left(\sum_m e^{-im(k_x-k_x')}\right)\left(\sum_l e^{-ilq(k_y-k_y')} \right) \times \\ \sum_{j} \alpha_j e^{-ijk_y}e^{ijk_y'} e^{-i2\pi\phi j} \; c_{k_x,k_y,j}^{\dagger} c_{k_x',k_y',j} \\
    = \cancel{\frac{q}{N_x N_y}} \sum_{k_x,k_y}\sum_{k_x',k_y'} e^{-ik_x}
    \cancel{N_x}\delta(k_x,k_x')\cancel{\frac{N_y}{q}}\delta(k_y,k_y') \sum_{j} \alpha_j \cancel{e^{-ijk_y}e^{ijk_y'}} e^{-i2\pi\phi j} c_{k_x,k_y,j}^{\dagger}  c_{k_x',k_y',j} \\
    = \sum_{k_x=-\pi}^{\pi}\sum_{k_y=-\pi/q}^{\pi/q} \; \sum_{j=0}^{q-1} \alpha_j e^{-ik_x} e^{-i2\pi\phi j} c_{k_x,k_y,j}^{\dagger} c_{k_x,k_y,j}.
\end{gather*}
\noindent
The sum of this quantity plus its Hermitian conjugate, all multiplied by $-t_a$, thus gives that
\begin{gather*}
    H_x = \sum_{k_x=-\pi}^{\pi}\sum_{k_y=-\pi/q}^{\pi/q} \; \sum_{j=0}^{q-1} \left[-t_a \alpha_j 
      e^{-i(k_x+2\pi\phi j)}  c_{k_x,k_y,j}^{\dagger} c_{k_x,k_y,j} + \textnormal{h.c.}\right]
\end{gather*}
\noindent
or, simplifying, 
$$
    H_x = \sum_{k_x=-\pi}^{\pi}\sum_{k_y=-\pi/q}^{\pi/q} \; \sum_{j=0}^{q-1} -2t_a \alpha_j \cos(k_x+2\pi\phi j)  
      c_{k_x,k_y,j}^{\dagger} c_{k_x,k_y,j}.
$$
\noindent
\hspace{\parindent}
To obtain a similar expression for $H_y$, note that
$$
    c_{m,n+1}^{\dagger}c_{m,n} = \frac{q}{N_x N_y} \sum_{k_x,k_y}\sum_{k_x',k_y'} e^{-i[mk_x+(lq+j+1)k_y]} e^{i[mk_x'+(lq+j)k_y']} c_{k_x,k_y,j+1}^{\dagger} c_{k_x',k_y',j}
$$
and
\begin{gather*}
    \sum_{m,n}\sqrt{\alpha_{n+1}} \sqrt{\alpha_n} c_{m,n+1}^{\dagger}c_{m,n} = \frac{q}{N_x N_y} \sum_{k_x,k_y}\sum_{k_x',k_y'} 
    \left(\sum_m e^{-im(k_x-k_x')}\right)\left(\sum_l e^{-ilq(k_y-k_y')} \right) \times \\ \sum_{j=0}^{q-1} \sqrt{\alpha_{j+1}}\sqrt{\alpha_j} e^{-i(j+1)k_y}e^{ijk_y'} c_{k_x,k_y,j+1}^{\dagger} c_{k_x',k_y',j} \\
    = \sum_{k_x=-\pi}^{\pi}\sum_{k_y=-\pi/q}^{\pi/q} \; \sum_{j=0}^{q-1} 
     \sqrt{\alpha_{j+1}}\sqrt{\alpha_j} e^{-ik_y}c_{k_x,k_y,j+1}^{\dagger} c_{k_x,k_y,j}
\end{gather*}
\noindent
which gives that
$$
    H_y = \sum_{k_x=-\pi}^{\pi}\sum_{k_y=-\pi/q}^{\pi/q} \; \sum_{j=0}^{q-1} -t_b \left[
     \sqrt{\alpha_{j+1}}\sqrt{\alpha_j} e^{-ik_y}c_{k_x,k_y,j+1}^{\dagger} c_{k_x,k_y,j} + \sqrt{\alpha_{j-1}}\sqrt{\alpha_{j}} e^{ik_y} c_{k_x,k_y,j-1}^{\dagger} c_{k_x,k_y,j}\right].
$$
\noindent
Overall then, the reciprocal space Hamiltonian (when $t_a=t_b=t$) is
\begin{gather*}
    H = -t\sum_{k_x=-\pi}^{\pi}\sum_{k_y=-\pi/q}^{\pi/q} \; \sum_{j=0}^{q-1} -2 \alpha_j \cos(k_x+2\pi\phi j)  
      c_{k_x,k_y,j}^{\dagger} c_{k_x,k_y,j} \\ + 
      \sqrt{\alpha_{j+1}}\sqrt{\alpha_j}e^{-ik_y} c_{k_x,k_y,j+1}^{\dagger} c_{k_x,k_y,j} + e^{ik_y} \sqrt{\alpha_{j-1}}\sqrt{\alpha_j}c_{k_x,k_y,j-1}^{\dagger} c_{k_x,k_y,j}.
\end{gather*}

Lastly, the correspondence of the potential energy term in Eqn. \ref{eqn:gutz_ham_r} and the reciprocal space definition provided in the manuscript can be checked combinatorically; the sum of effective Coulomb contributions over the real-space lattice points must equal the sum of effective Coulomb contributions over both the reciprocal lattice points and unit cell plaquettes.

\section{Derivation of Velocity Operators}

One may derive the form of the velocity operators starting from the following commutation relation \cite{Dutta_jap2012},
\begin{gather*}
    \hat{v} = \frac{1}{i\hbar} \left[\hat{r},\hat{H}\right] = \frac{1}{i\hbar} \left(\hat{r}\hat{H}-\hat{H}\hat{r}\right) = -\frac{i}{\hbar}\sum_{i,j,k} \left(c_{i}^{\dagger} c_{i} \cdot c_{j}^{\dagger} H_{jk} c_{k} - c_{j}^{\dagger} H_{jk} c_{k} \cdot c_{i}^{\dagger}  c_{i}\right) \\
    = -\frac{i}{\hbar}\sum_{i,j,k} \left(c_{i}^{\dagger}  (\delta_{ij}) H_{jk} c_{k} - c_{j}^{\dagger} H_{jk} (\delta_{ki})  c_{i}\right) = -\frac{i}{\hbar}\sum_{i,j,k} \left(c_{i}^{\dagger} H_{ik} c_{k} - c_{j}^{\dagger} H_{ji}  c_{i}\right) \\ = -\frac{i}{\hbar}\sum_{i,j,k} \sum_{\ell} (c_{i}^{\dagger}  H_{ik} \cancelto{\delta_{k \ell}}{c_{k} c_{\ell}^{\dagger}} c_{\ell} - c_{\ell}^{\dagger} \cancelto{\delta_{\ell j}}{c_{\ell}c_{j}^{\dagger}} H_{ji}  c_{i}) = -\frac{i}{\hbar}\sum_{i,\ell} \left(c_{i}^{\dagger}  H_{il} c_{\ell} - c_{\ell}^{\dagger} H_{li}  c_{i}\right)
    = -\frac{i}{\hbar}\sum_{i,\ell} \left(c_{i}^{\dagger}  H_{il} c_{\ell} - c_{\ell}^{\dagger} H_{il}^{*}  c_{i}\right)
\end{gather*}
which gives a general tight-binding expression for the velocity operator,
\begin{equation}
    \hat{v} = \frac{i}{\hbar}\sum_{i,\ell} \left(c_{\ell}^{\dagger} c_{i}H_{il}^{*} - c_{i}^{\dagger} c_{\ell}H_{il} \right).
\end{equation}
Substituting the Hamiltonian of Equation \ref{eqn:gutz_ham_r} into the general expression for the velocity operator and letting $\ell \rightarrow \left\lbrace m,n \right\rbrace$, $i \rightarrow \left\lbrace m+1,n \right\rbrace$, the Gutzwiller-renormalized velocity operator in $x$ is
\begin{equation}
    \hat{v}_x = -\frac{it_a}{\hbar}\sum_{m,n} \sqrt{\alpha_{m,n}} \sqrt{\alpha_{m+1,n}} \left(c_{m,n}^{\dagger} c_{m+1,n}e^{i2\pi \Phi n} - c_{m+1,n}^{\dagger} c_{m,n}e^{-i2\pi \Phi n} \right).
\end{equation}
Similarly, letting $\ell \rightarrow \left\lbrace m,n \right\rbrace$ and $i \rightarrow \left\lbrace m,n+1 \right\rbrace$, the velocity operator in $y$ is
\begin{equation}
    \hat{v}_y = -\frac{it_b}{\hbar}\sum_{m,n} \sqrt{\alpha_{m,n}} \sqrt{\alpha_{m,n+1}} \left(c_{m,n}^{\dagger} c_{m,n+1} - c_{m,n+1}^{\dagger} c_{m,n} \right).
\end{equation}

\section{Effect of Site Renormalization Parameters on Band Structure}

\begin{figure*}[hb!]
\renewcommand{\thefigure}{S1}
\includegraphics[width=6.5in]{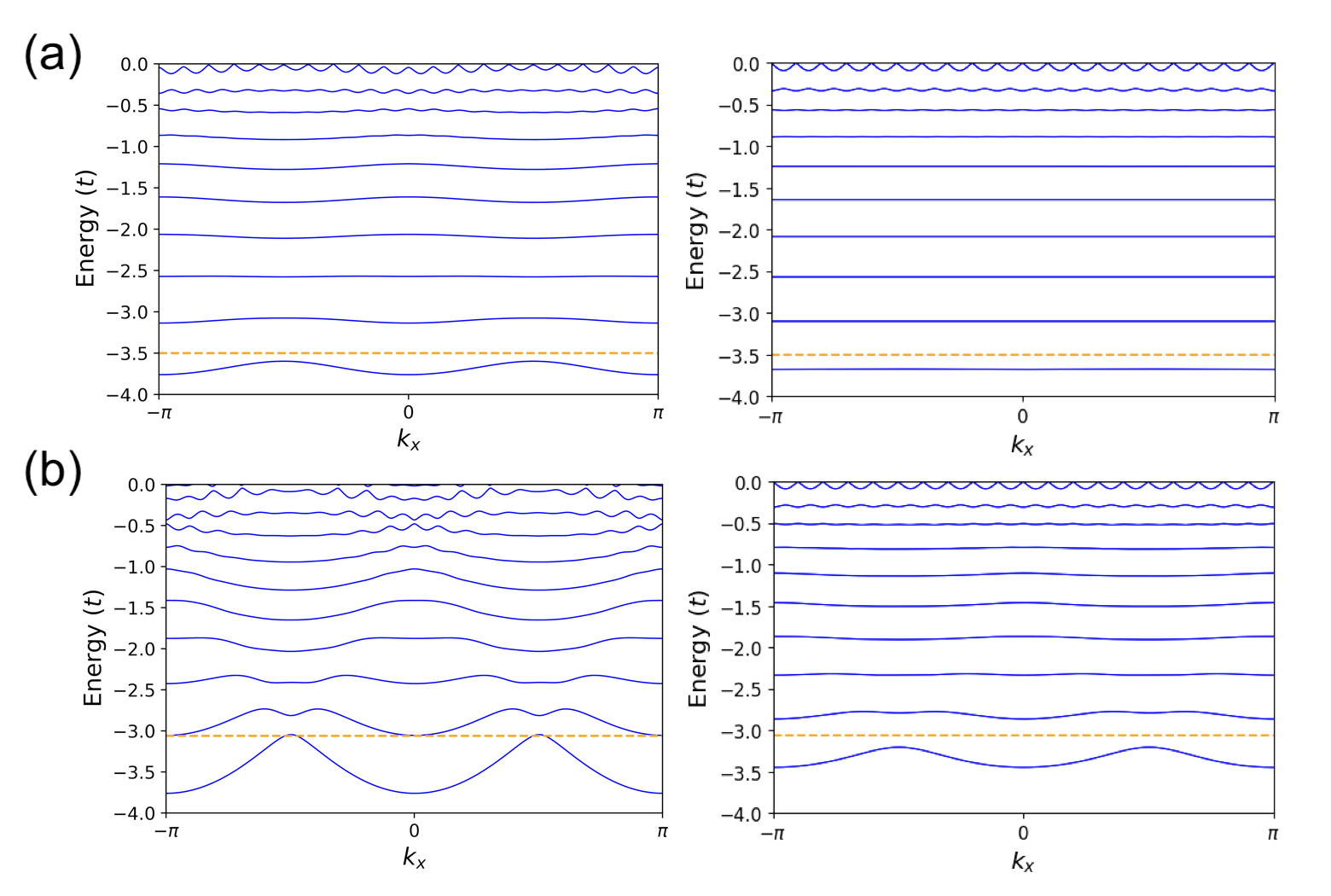}
%\vspace*{-15mm}
\caption{Plots of the band structures of (a) weakly correlated ($U=1t$) and (b) strongly correlated ($U=5t$) magnetized square lattice models. The bands on the left include the optimal site renormalization parameters $\lambda_{j,s}$ while those on the right do not.}
\label{fig:supp_lambdas}
\end{figure*}

The renormalized band structures in the presence and absence of site renormalization parameters $\lambda_{j,s}$ are shown in Figure \ref{fig:supp_lambdas}. In the presence of both weak and strong correlation ($U = 1t$ and $U=5t$), band bending is enhanced by the presence of the site renormalization parameter. In the absence of this parameter, the bands bend to a much smaller extent; in this case the effect is not large enough to close the mobility gap between the first and second LLs. Thus, the degradation of the first Landau level depends on both hopping renormalization and effective charge reordering, quantified by the site renormalization energies.

%\bibliography{supplement}

%\end{document}